\documentclass[aps,prl,twocolumn,superscriptaddress,preprintnumbers,showpacs]{revtex4}
\usepackage{graphics}
\def\eq#1\en{\begin{equation} #1 \end{equation}}

\def\eqa#1\ena{\begin{eqnarray} #1 \end{eqnarray}}

\usepackage{graphicx,color}

\begin{document}

\title{Skyrme model predictions for the ${\bf 27}_{J=3/2}$ mass spectrum\\ and 
the ${\bf 27}_{3/2}$--$\overline{\bf 10}$ mass splittings}

\author{G. Duplan\v{c}i\'{c}}
\affiliation{Theoretical Physics Division, Rudjer Bo\v skovi\' c Institute, 
Zagreb, Croatia}
\author{H. Pa\v{s}agi\'{c}}
\affiliation{Faculty of Transport and Traffic Engineering, University of Zagreb, \\
P.O. Box 195, 10000 Zagreb, Croatia} 
\author{J.Trampeti\'{c}}
\affiliation{Theoretical Physics Division, Rudjer Bo\v skovi\' c Institute, 
Zagreb, Croatia}
\affiliation{Theory Division, CERN, CH-1211 Geneva 23, Switzerland}
\affiliation{Theoretische Physik, Universit\"{a}t M\"{u}nchen, Theresienstr. 37, 80333 M\"{u}nchen, Germany}

\date{\today}

\begin{abstract}
The ${\bf 27}_{J=3/2}$-plet mass spectrum and the ${\bf 27}_{3/2}$--$\overline{\bf 10}$ mass splittings   
are computed in the framework of the minimal SU(3)$_f$ extended Skyrme model.
As functions of the Skyrme charge $e$ and the SU(3)$_f$ symmetry breaking parameters
the predictions are presented in tabular form.
The predicted mass splitting ${\bf 27}_{3/2}$--$\overline{\bf 10}$
is the smallest among all SU(3)$_f$ baryonic multiplets, confirming earlier findings. 
\end{abstract}


\maketitle

The discovery \cite{penta1} 
of the exotic baryon $\Theta^+$, with strangeness +1 and probable spin 1/2, 
recently supported by the observations of $\Theta^+$ in various experiments 
\cite{penta2,penta3,penta4,penta5},
and the discovery \cite{penta6} of the exotic isospin 3/2 baryon 
with strangeness -2, $ \Xi^{--}_{3/2}$, have produced 
huge excitement in the high energy physics community.

The $\Theta^+$-baryon mass was  
successfully predicted in the ``model-independent'' way for the first time in Ref. \cite{MP}.
However, it was the prediction of the narrow width of $\Theta^+$ in the chiral 
quark-soliton model of Ref. \cite{dia1} that stimulated experimental searches.
To estimate baryon multiplets (${\bf 8},\;{\bf 10},\;{\overline{\bf 10}},\;{\bf 27}, \; {\rm etc.} $) 
mass spectra, relevant mass differences and other baryon properties,
various authors employed different types of methods and models 
\cite{BD,Manohar,Chemtob,HW,wei,wei1,WK,K,Glo,
Itz,Kar,Zhu,Ger,lattice,Hua,JW,DP,WM,faber,Ell,DT,JM,cohen,Weigel,Rossi}. 

The main aim of this Brief Report is the application of 
the minimal SU(3)$_f$ extended Skyrme model \cite{wei} in an attempt to predict 
the ${\bf 27}_{3/2}$--$\overline{\bf 10}$ mass splitting and the ${\bf 27}_{3/2}$-plet mass spectrum.
The minimally extended Skyrme model uses only one free parameter, the Skyrme charge $e$,
and only flavor symmetry breaking (SB) term proportional to $\lambda_8$ in the action
$
{\cal L} = {\cal L}_{\sigma} + {\cal L}_{\rm Sk} + {\cal L}_{\rm WZ} + {\cal L}_{\rm SB} ,
$
where  ${\cal L}_{\sigma}$, ${\cal L}_{\rm Sk}$, ${\cal L}_{\rm WZ}$, and ${\cal L}_{\rm SB}$ denote 
the $\sigma$-model, Skyrme, Wess--Zumino and SB terms 
\cite{sky,tho,witt,adk,gua,wess,yab}, respectively.
For the profile function in ${\cal L}$ we use the arctan ansatz \cite{dia}
which makes possible to evaluate relevant overlap integrals analytically.
The classical soliton mass ${\cal M}_{\rm csol}$ 
receives too large a value producing an unrealistic baryonic mass spectrum. 
We are using it only
to obtain the dimensionless size of the skyrmion $x_0$ by minimizing ${\cal M}_{\rm csol}(x_0)$.
The dimensionless size of the skyrmion $x_0$  
includes the dynamics of SB effects, which takes place within the skyrmion. 
It follows that \cite{DT}: 
\begin{eqnarray}
x^2_0 = \frac{15}{8} \left[ 1 + \frac{6\beta'}{f^2_{\pi}} + 
\sqrt {\left(1+ \frac{6\beta'}{f^2_{\pi}}\right)^2 + 
\frac{30\delta'}{e^2 f^4_{\pi}}} \; \right]^{-1},
\label{2} 
\end{eqnarray}
where the SB parameters ${\hat x}, {\beta}', {\delta}'$ are given by \cite{wei}:
\begin{eqnarray}
 {\hat x}&=&\frac{2m^2_K f^2_K}{m^2_{\pi}f^2_{\pi}} -1, \;\;\beta' =\frac{f^2_K - f^2_{\pi}}{4(1-{\hat x})}, 
\nonumber\\
\delta' &=& \frac{m^2_{\pi}f^2_{\pi}}{4} =\frac{m^2_K f^2_K}{2(1+{\hat x})}.
\label{3}
\end{eqnarray}
The symmetry breaker ${\hat x}$ was constructed systematically from the QCD mass term in the case of SU(3)$_f$.
The $\delta'$ term is required to split pseudoscalar meson masses, while the $\beta'$ term is required
to split pseudoscalar decay constants (for details, see Ref. \cite{wei}).

To obtain the ${\bf 27}_{3/2}$--$\overline{\bf 10}$ mass splittings 
and the {\bf 27}$_{3/2}$ mass spectrum, the following 
definition of the mass formulas are used:
\begin{eqnarray}
{\rm M}^{\overline{\bf 10}}_B(x_0) \hspace{-1mm}&=& {\cal M}^{\bf 8} +\frac{3}{2\lambda_s (x_0)}- 
\frac{\gamma(x_0) }{2} \;\delta^{\overline{\bf 10}}_B ,
\label{4}\\
{\rm M}^{\bf 27}_B(x_0) \hspace{-1mm}&=& {\cal M}^{\bf 8} 
+\frac{3}{2\lambda_c (x_0)} +\frac{1}{\lambda_s (x_0)}- \frac{\gamma(x_0)}{2} \;\delta^{\bf 27}_{B}.
\label{5}
\end{eqnarray}
Here the experimental octet mean mass 
${\cal M}^{\bf 8}=\frac{1}{8}\sum_{B=1}^{\bf 8}{\rm M}^{\bf 8}_B=1151$ MeV
was used instead of 
${\cal M}^{\bf 8}={\cal M}_{\rm csol}(x_0)+\frac{3}{2\lambda_c (x_0)}$. 
From experiment it also follows that the decuplet mean mass
${\cal M}^{\bf 10}=\frac{1}{10}\sum^{10}_{B=1}{\rm M}^{\bf 10}_B =1382$ MeV \cite{rpp}.  
The splitting constants $\delta^{\overline{\bf 10}}_B$ and 
$\delta^{\bf 27}_B$ are given in \cite{Ell} and in Table 1. of Ref. \cite{WM}, respectively. 
The moment of inertia $\lambda_c$ for rotation in coordinate space, 
and the moment of inertia $\lambda_s$ for flavor rotations in the direction of the strange degrees of freedom,
except for the eighth direction, 
and the symmetry breaking quantity $\gamma$,  
[the coefficient in the SB 
piece ${\cal L}_{\rm SB} = -\frac{1}{2} \gamma (1-D_{88})$ of a total collective Lagrangian ${\cal L}$], 
are given in Ref. \cite{DT}. 

From (\ref{4}) and (\ref{5}) the ${\bf 27}_{3/2}$--$\overline{\bf 10}$ 
mean mass splitting ${\Delta}^{\overline{\bf 10}}_{\bf 27}$ is given by 
\begin{eqnarray}
{\Delta}^{\overline{\bf 10}}_{\bf 27}&\equiv&
{\cal M}^{\bf 27}_{3/2} - {\cal M}^{\overline{\bf 10}}=
 \label{6}\\
&=&\frac{1}{2}\left[\frac{3}{\lambda_c (x_0)}-\frac{1}{\lambda_s (x_0)}\right]
\equiv {\Delta}^{\bf 10}_{\bf 8}-\frac{1}{3}{\Delta}^{\overline{\bf 10}}_{\bf 8},
\nonumber
\end{eqnarray}
where ${\Delta}^{\overline{\bf 10}}_{\bf 27}$ is also expressed in terms 
of the decuplet--octet ${\Delta}^{\bf 10}_{\bf 8}$ and the antidecuplet--octet 
${\Delta}^{\overline{\bf 10}}_{\bf 8}$ mean mass splittings, \cite{DT}.
In the computations of the mean masses ${\cal M}^{\overline{\bf 10}}$ 
and ${\cal M}^{\bf 27}_{3/2}$
the sum of $D_{88}$ diagonal elements over all components of irreducible representations
cancels out because of the properties of the SU(3) Clebsh-Gordan coefficients.

The mass splittings between the same quark flavor content baryons of ${\bf 27}_{3/2}$ and 
$\overline{\bf 10}$-plets are: 
\begin{eqnarray}
\delta_1
&=&{\rm M}^{\bf 27}_{3/2}(\Theta_1)-{\rm M}^{\overline{\bf 10}}(\Theta^+)
={\Delta}^{\overline{\bf 10}}_{\bf 27}+\frac{3}{56}\gamma(x_0),
\label{7}\\
\delta_2
&=&{\rm M}^{\bf 27}_{3/2}(N^*_{\frac{3}{2}})-{\rm M}^{\overline{\bf 10}}(N^*)
={\Delta}^{\overline{\bf 10}}_{\bf 27}+\frac{1}{224}\gamma(x_0),
\nonumber\\
\delta_3
&=&{\rm M}^{\bf 27}_{3/2}(N^*_{\frac{1}{2}})-{\rm M}^{\overline{\bf 10}}(N^*)
={\Delta}^{\overline{\bf 10}}_{\bf 27}+\frac{5}{112}\gamma(x_0),
\nonumber\\
\delta_4
&=&{\rm M}^{\bf 27}_{3/2}(\Sigma_2)-{\rm M}^{\overline{\bf 10}}(\Sigma)
={\Delta}^{\overline{\bf 10}}_{\bf 27}-\frac{5}{112}\gamma(x_0),
\nonumber\\
\delta_5
&=&{\rm M}^{\bf 27}_{3/2}(\Sigma_1)-{\rm M}^{\overline{\bf 10}}(\Sigma)
={\Delta}^{\overline{\bf 10}}_{\bf 27}+\frac{1}{112}\gamma(x_0),
\nonumber\\
\delta_6
&=&{\rm M}^{\bf 27}_{3/2}(\Lambda^*)-{\rm M}^{\overline{\bf 10}}(\Sigma)
={\Delta}^{\overline{\bf 10}}_{\bf 27}+\frac{1}{28}\gamma(x_0),
\nonumber\\
\delta_7
&=&{\rm M}^{\bf 27}_{3/2}(\Xi^*_{\frac{3}{2}})-{\rm M}^{\overline{\bf 10}}(\Xi_{\frac{3}{2}})
={\Delta}^{\overline{\bf 10}}_{\bf 27}-\frac{3}{112}\gamma(x_0),
\nonumber\\
\delta_8
&=&{\rm M}^{\bf 27}_{3/2}(\Xi^*_{\frac{1}{2}})-{\rm M}^{\overline{\bf 10}}(\Xi_{\frac{3}{2}})
={\Delta}^{\overline{\bf 10}}_{\bf 27}+\frac{3}{224}\gamma(x_0).
\nonumber
\end{eqnarray}
The $\Xi$ isoquartet and isodoublet from the $\bf 27$, spin 3/2, we mark as
$\Xi^*_{\frac{3}{2}}$ and $\Xi^*_{\frac{1}{2}}$, 
to distinguish them from the the $\Xi$ isoquartet and isodoublet 
from the $\overline{\bf 10}$, spin 1/2. We also mark the $\bf 27$-plet
isosinglet as $\Lambda^*$.

Considering the SB parameters (\ref{3}), at this point we introduce three 
different dynamical assumptions based on the SB part of the Lagrangian producing three
fits which will be used further in our numerical analysis:
\begin{eqnarray}
&({\rm i})& \hspace{2mm} m_\pi=m_K=0, \;\,f_\pi=f_K=93\;\, {\rm MeV}\; 
\nonumber\\
&& \hspace{2mm}\Longrightarrow \,{\hat x}=1,\;\, \beta' = \delta' = 0;
\nonumber\\
&({\rm ii})&\; m_\pi=138,\;\, m_K=495, \;\,f_\pi=f_K=93\,\; {\rm MeV}\; 
\nonumber\\
&&\hspace{2mm}\Longrightarrow \,{\hat x}=24.73,\,\;\beta'=0,\,\;\delta'=4.12\,\times 10^7\,{\rm MeV}^4;
\nonumber\\
&({\rm iii})& \,m_\pi=138,\,\; m_K=495, \,f_{\pi} =93,\,f_K =113\, {\rm MeV} 
\nonumber\\
&&\hspace{2mm}\Longrightarrow \,{\hat x}=36.97,\;\, \beta' = -28.6\;\,{\rm MeV}^2,
\nonumber\\
&&\hspace{2mm}\delta' = 4.12\,\times 10^7\,{\rm MeV}^4.
\label{8}
\end{eqnarray}
Switching off SU(3)$_f$ symmetry breaking, which corresponds to case (i), the  
absolute masses of each member of the multiplet become equal for each fixed $e$.
In the chiral limit, 
\begin{eqnarray}
x_0=\frac{\sqrt{15}}{4} \;\Rightarrow \;
{\Delta}^{\overline{\bf 10}}_{\bf 27}=\delta_{1,\ldots,8}
=\frac{52e^3f_{\pi}}{285\sqrt{30}\pi^2}.
\label{9}
\end{eqnarray}
For example, from (\ref{5}) and (\ref{9}) one would have ${\rm M}^{\bf 27}_{3/2}=1898$
MeV and ${\Delta}^{\overline{\bf 10}}_{\bf 27}=32.6$ MeV, for $e=4.7$ .

The mass splittings (\ref{6}) and (\ref{7})
as functions of two different dynamical assumptions, (ii,iii), and the Skyrme charge $e$ are
given in Table \ref{t:tab1}.
We have chosen four values of the Skyrme charge $e=3.4,\;4.2,\;4.4,\;4.7$
because in the minimal approach they give the best fit for
the nucleon axial coupling constant $g_A = 1.25$ \cite{dppt},
the mass splitting $({\Delta}^{\bf 10}_{\bf 8})_{\rm exp}=231$ MeV, 
and the penta-quark masses $\rm M^{\rm exp}_{\Theta^+} = 1540$ MeV
and ${\rm M}^{\rm exp}_{\Xi^{--}_{3/2}}$=1861 MeV, respectively.
\renewcommand{\arraystretch}{1.4}
\begin{table}
\caption{The ${\bf 27}_{3/2}-{\overline{\bf 10}}$ mass splittings (MeV) 
as functions of the Skyrme charge $e$ and for fits (ii), (iii).}
\begin{center}
\begin{tabular}{|c|ccccc|ccccc|}
\hline
${\rm Fit}$ & $$ & $$& $(\rm ii)$ & $$ & $$ & $$ & $$ & $(\rm iii)$ & $$ & $$ \\
\hline
$ e $ & $3.2$ & $3.4$ & $4.2$ & $4.4$ & $4.7$ & $3.2$ & $3.4$ & $4.2$ & $4.4$ & $4.7$ \\
\hline \hline
${\Delta}^{\bf 10}_{\bf 8}$ & $ 110$ & $ 129$ & $ 229$ & $ 260$ & $ 312$ & $ 109$ 
& $ 128$ & $ 227$ & $ 257$ & $ 309$\\

${\Delta}^{\overline{\bf 10}}_{\bf 8}$ & $302$ & $354$ & $621$ & $704$ & $843$ & $ 233$
& $273$ & $474$ & $ 536$ & $ 641$\\
 
${\Delta}^{\overline{\bf 10}}_{\bf 27}$ & $ 9$ & $11 $ & $ 22$ & $ 25$ & $ 31$ 
& $31$ & $ 37$ & $ 69$ & $79$ & $ 95$\\
 \hline
$\delta_1$ & $ 99$ & $89 $ & $ 67$ & $ 66 $ & $ 65 $ & $ 179 $ & 
$ 165 $ & $ 146 $ & $ 148 $ & $ 154 $\\

$\delta_2$ & $ 17$ & $18 $ & $ 26$ & $ 29 $ & $34 $ &  $ 44 $ 
& $ 48 $ & $ 75 $ & $ 84 $ & $ 100 $\\

$\delta_3$ & $ 84$ & $76 $ & $ 60$ & $ 59 $ & $ 59 $ & $ 154 $ 
& $ 144 $ & $ 133 $ & $ 136 $ & $ 144 $\\

$\delta_4$ & $ -66$ & $-53 $ & $ -16$ & $ -8 $ & $ 3 $ & $ -91 $ & 
$ -69 $ & $ 4 $ & $ 21 $ & $ 46 $\\

$\delta_5$ & $ 24$ & $24 $ & $ 30$ & $ 32 $ & $ 37 $ & $ 56 $ & 
$ 59 $ & $ 82 $ & $ 90 $ & $ 105 $\\
 
$\delta_6$ & $ 69$ & $ 63$ & $ 52$ & $ 52$ & $ 54$ & $ 130$ 
& $ 123$ & $120 $ & $ 125$ & $ 134$\\

$\delta_7$ & $ -36$ & $-27 $ & $ -1$ & $ 5 $ & $ 14 $ & $ -42 $ 
& $-27 $ & $30 $ & $44 $ & $66 $\\

$\delta_8$ & $ 32$ & $31 $ & $33$ & $ 36 $ & $ 40 $ & $ 68 $ 
& $ 69 $ & $ 88 $ & $ 96 $ & $ 110 $\\
\hline 
\end{tabular}
\label{t:tab1}
\end{center}
\end{table}
\renewcommand{\arraystretch}{1}

Assuming equal spacing for antidecuplets, 
from the recent experimental data ($\rm M^{\rm exp}_{\Theta^+}=1540$ MeV 
and $\rm M^{\rm exp}_{\Xi^{--}_{3/2}}=1861$ MeV), 
in Ref. \cite{DT} we have found the following
masses of antidecuplets $\rm M_{N^*}=1647$ MeV, 
$\rm M_{\Sigma_{\overline{10}}} =1754$ MeV, the mean mass  
${\cal M}^{\overline{\bf 10}}=\frac{1}{10}\sum^{10}_{B=1}{\rm M}^{\overline{\bf 10}}_B =1754$ MeV
and the mass difference ${\Delta}^{\overline{\bf 10}}_{\bf 8}$= 603 MeV. 
Taking 603 MeV, bonafide, as an ``experimental'' estimates for ${\Delta}^{\overline{\bf 10}}_{\bf 8}$,
together with $({\Delta}^{\bf 10}_{\bf 8})_{\rm exp}=231$ MeV,
via Eq. (\ref{6}), we estimate ${\Delta}^{\overline{\bf 10}}_{\bf 27}=30$ MeV.
It turns out from Table \ref{t:tab1} that only $e\,\simeq\,3.2$, 
in the most realistic case (iii), could account for the small value of 
${\Delta}^{\overline{\bf 10}}_{\bf 27}$.
However, $e=3.2$ gives too small values for ${\Delta}^{\bf 10}_{\bf 8}$   
and ${\Delta}^{\overline{\bf 10}}_{\bf 8}$.

Using 1754 MeV for the $\overline{\bf 10}$-plet mean mass and 
the predicted range for the mean mass splitting 
$30\,\leq \,{\Delta}^{\overline{\bf 10}}_{\bf 27}\,\leq \,95$ MeV, we find the
range for the ${\bf 27}_{3/2}$-plet mean mass as $1784\,\leq {\cal M}^{\bf 27}_{3/2}\,\leq 1849$ MeV, which is 
approximately placed into the center of the ${\bf 27}_{3/2}$-plet mass spectrum displayed in Fig. 4 of Ref. \cite{WK}
(for A and B fits), and in Fig. 4 of Ref. \cite{Ell}.
Careful inspection of the results for the ${\bf 27}_{3/2}$-plet mass spectrum from Fig. 4
of Ref. \cite{WK} shows approximate agreement with our results, $\delta_1,...,\delta_8$,
for $4.2 \,\leq \,e\,\leq 4.7$ fit (iii), presented in Table \ref{t:tab1}.

\renewcommand{\arraystretch}{1.4}
\begin{table}
\caption{The {\bf 27}$_{3/2}$ mass spectrum (MeV) 
as functions of the Skyrme charge $e$ and for fits (ii), (iii).}
\begin{center}
\begin{tabular}{|c|ccccc|ccccc|}
\hline
${\rm Fit}$ & $$ & $$ & $({\rm ii})$ & $$ & $$ & $$ & $$ & $ ({\rm iii}) $ & $$ & $$ \\
\hline
$ e $ & $3.2$ & $3.4$ & $4.2$ & $4.4$ & $4.7$ & $3.2$ & $3.4$ & $4.2$ & $4.4$ & $4.7$ \\
\hline \hline
$\Theta_1$ & $ 1343$ & $ 1413$ & $ 1734$ & $ 1827 $ & $ 1980 $ & $ 1219 $ & $ 1290 $ & $ 1590 $ & 
$ 1674 $ & $ 1808 $ \\

$\rm N^*_{\frac{3}{2}}$ & $ 1365$ & $1433 $ & $ 1745$ & $ 1837 $ & $ 1988 $ & $ 1256 $ &  
$ 1322 $ & $ 1610 $ & $ 1691 $ & $ 1823 $ \\
 
$\Sigma_2$ & $ 1388$ & $1452 $ & $ 1756$ & $ 1847 $ & $ 1997 $ & $ 1293 $ & $1354 $ &  
$ 1629 $ & $ 1708 $ & $ 1838 $  \\
 
$\rm N^*_{\frac{1}{2}}$ & $ 1433$ & $1491 $ & $ 1779$ & $ 1867 $ & $ 2014 $ & $ 1367 $ & 
$ 1418 $ &  $ 1668 $ & $ 1743 $ & $ 1867 $ \\

$\Sigma_1$ & $ 1478$ & $1529 $ & $ 1802$ & $ 1887 $ & $ 2031 $ &  $ 1440 $ & $ 1482 $ & 
$ 1706 $ & $ 1778 $ & $ 1897 $ \\

$\Lambda^*$ & $ 1522$ & $1568 $ & $ 1824$ & $ 1908 $ & $ 2048 $ & $ 1514 $ & $ 1546 $ & 
$ 1745 $ & $ 1812 $ & $ 1926 $ \\

$\Xi^*_{\frac{3}{2}}$ & $ 1522$ & $1568 $ & $ 1824$ & $ 1908 $ & $ 2048 $ & $ 1514 $ & 
$ 1546 $ &  $ 1745 $ & $ 1812 $ & $ 1926 $ \\

$\Xi^*_{\frac{1}{2}}$ & $ 1590$ & $1626 $ & $ 1858$ & $ 1938 $ & $ 2073 $ & $ 1624 $ & 
$ 1642 $ &  $ 1803 $ & $ 1864 $ & $ 1970 $ \\
 
$\Omega_1$ & $ 1657$ & $1684 $ & $ 1892$ & $ 1968 $ & $ 2099 $ & $ 1735 $ & $ 1738 $ &  
$ 1861 $ & $ 1916 $ & $ 2014 $ \\
\hline 
\end{tabular}
\label{t:tab2}
\end{center}
\end{table}
\renewcommand{\arraystretch}{1}
Comparing the pure Skyrme model
prediction of Ref. \cite{WK} (fits A and B in Figure 4) with our results from Table \ref{t:tab2},
we have found that our case (iii) with $4.3\,\leq\,e\,\leq \,4.7$ supports fit B, and
for $4.4\,\leq\,e\,\leq \,4.6$ agrees nicely with fit A. Both fits A and B from \cite{WK}
lie between $4.0\,\leq\,e\,\leq \,4.6$ for case (ii).
Case (iii) with $4.2\,\leq\,e\,\leq \,4.7$ also
supports the results presented in Table 1. of Ref. \cite{WM}.
From Table \ref{t:tab2} we conclude that the best fit for 
the {\bf 27}$_{3/2}$ baryon mass spectrum,
as a function of $e$ and for $f_K\not=f_{\pi}$, would lie between $e\simeq 4.2$ and $e\simeq 4.7$,
just like that for the octet, decuplet and anti-decuplet mass spectra \cite{DT}.
In Table \ref{t:tab2} the masses of $\Lambda^*$ and $\Xi^*_{\frac{3}{2}}$ are equal owing to the absence
of anomalous moments of inertia \cite{MP,HW} in the model used.
Note, however, that the anomalous moments of inertia 
contributions are estimated to be at best $\sim$ 1 \% for the $\Xi^*_{\frac{3}{2}}$ mass
\cite{WM,Ell}, for example. 

Next we comment on possible effects coming from the mixing between exotic rotational excitations
and vibrational (or radial) excitations \cite{cohen,Weigel} in the minimal SU(3)$_f$ extended Skyrme model. 
Let us note that, in the case of the ${\bf 27}$-plet, states with $Y=\pm 2$ and $Y=+1$, $I=3/2$
do not mix with neither ${\bf 8}$, ${\bf 10}$ or $\overline{\bf 10}$, nor with their
vibrational excitations. They will have vibrational excitations themselfs, but, as results of Ref. 
\cite{Weigel} indicate, such vibrations are expected to have minor influence on ``base'' states. 
Therefore for these states our predictions are correct within the approximations made, i.e. 
by neglecting  $1/N_c$ corrections to ${\cal L}_{\rm SB}$. All other states will be subject to mixing.
However, their masses, given in Table \ref{t:tab2}, represents the predictions under no mixing assumption.   
Considering the question of the decay width calculations, 
the Skyrme model is too crude to give reliable predictions for the widths \cite{WM,Ell}.
Here the $1/N_c$ corrections, missing in the present approach, are of primary importance \cite{dia1}.

For the simplest version of the total Lagrangian,
the results given in Tables \ref{t:tab1} and \ref{t:tab2}
do agree well with the other Skyrme model based estimates \cite{MP,dia1,wei1,WK,K,WM,Ell}. 
In particular, our approach is similar to \cite{WK,K}. 

As has been discussed in \cite{dppt}, although
the symmetry breaking effects are generally very important, 
the main effect comes from the $D_{88}$ term
confirming the results of \cite{WK,K,Ell,Weigel}.
In our approach, in the language of \cite{Weigel}, the reduction of 
the influence of the so called ${\overline s}s$ cloud was taken into account 
by inclusion of the SB term $(1-D_{88})$ in mass formulae (\ref{4}) and (\ref{5}). 

It is clear from Table \ref{t:tab1} that for fixed $e$ the difference between fits (ii) and (iii) 
is crucial for the correct description of the mass splittings (\ref{6}) and (\ref{7}). 
For small mass splittings the contribution of the term proportional to $(f^2_K-f^2_{\pi})$ in
the Lagrangian $\cal L$ plays a major role.

The ${\bf 27}_{3/2}$--$\overline{\bf 10}$ 
mass splittings are the quantities whose 
measured values, together with measurements of the decay modes branching ratios,  
would determine the spins, 3/2 or 1/2, of observed objects, like $\Xi^{--}_{3/2}$, 
thus placing it into the right SU(3)$_f$ representation.
We do expect that experimental analysis, considering other members of the 
$\overline{\bf 10}$  and ${\bf 27}_{3/2}$-plets, should also be performed. 

We hope that the present calculation, taken together with 
the analogous calculation in \cite{MP,wei,WK,K,WM,faber,Ell,DT} will
contribute to the understanding of the overall picture of 
the baryonic mass spectrum and mass splittings in the Skyrme model,
as well as to further computations of other nonperturbative, dimension-6 operator matrix elements between
different baryon states \cite{dppt,tra}.

Since the splittings (\ref{6},\ref{7}) represent 
the smallest splittings among splittings between 
the SU(3)$_f$ multiplets ${\bf 8}$, ${\bf 10}$, $\overline{\bf 10}$, ${\bf 27}$, ${\bf 35}$ and $\overline{\bf 35}$
we would urge our colleagues to continue experimental analysis of penta-quark spectral and decay modes  
and find the penta-quark members of the ${\bf 27}_{3/2}$-plet which would mix with or 
lie just above the penta-quark family of the $\overline{\bf 10}$-plet.

We would like to thank K. Kadija for  helpful discussions.
One of us (JT) would like to thank M. Praszalowicz for a careful reading of the manuscript 
and J. Wess for many stimulating 
discussions and Theoretische Physik, Universit\" at
M\" unchen and Theory Division CERN, where part of this work was done, for hospitality.
This work was supported by the Ministry of Science and Technology of the Republic of Croatia under Contract 0098002.

\end{document}